\documentclass[10pt]{article}

\usepackage{amsmath,amssymb,epsfig,color,bbold}
\usepackage{feynmp-auto}
\usepackage[hidelinks]{hyperref}
\usepackage{caption}
\usepackage{subcaption}
\usepackage{cancel}
\usepackage{soul}
\usepackage{slashbox}
\usepackage{physics}
\usepackage{psfrag}
\usepackage{graphicx,xspace}
\usepackage{bm}

\usepackage[affil-it, auth-sc]{authblk}
\usepackage{lineno}
\usepackage[sort&compress,numbers]{natbib}
\usepackage{doi}
\usepackage{eso-pic}
\usepackage[capitalise]{cleveref}

\newcommand{\be}{\begin{equation}}  
\newcommand{\ee}{\end{equation}} 
\def\slash#1{#1\!\!\!/\!\,\,}  
\newcommand{\nl}{\nonumber \\ }

\long\def\symbolfootnote[#1]#2{\begingroup%
\def\thefootnote{\fnsymbol{footnote}}\footnote[#1]{#2}\endgroup}

\def\dd{\mathrm{d}} 

\def\deltaE{\vb*{\varepsilon}_\gamma }
\def\me{m}

\newcommand{\iu}{{\rm i}}

\allowdisplaybreaks

\begin{document}

\begin{fmffile}{fmfnotes} 
\fmfcmd{%
vardef middir(expr p,ang) = dir(angle direction length(p)/2 of p + ang) enddef;
style_def arrow_left expr p = shrink(.7); cfill(arrow p shifted(4thick*middir(p,90))); endshrink enddef;
style_def arrow_left_more expr p = shrink(.7); cfill(arrow p shifted(6thick*middir(p,90))); endshrink enddef;
style_def arrow_right expr p = shrink(.7); cfill(arrow p shifted(4thick*middir(p,-90))); endshrink enddef;}

\fmfset{wiggly_len}{3mm}
\fmfset{wiggly_slope}{75}
\fmfset{curly_len}{2mm}
\fmfset{zigzag_len}{2mm}

\fmfset{arrow_ang}{15}
\fmfset{arrow_len}{2.5mm}
\fmfset{decor_size}{3mm}

\fmfstraight

\AddToShipoutPictureFG*{\AtPageUpperLeft{\put(-60,-75){\makebox[\paperwidth][r]{FERMILAB-PUB-25-0028-T}}}}
\AddToShipoutPictureFG*{\AtPageUpperLeft{\put(-60,-60){\makebox[\paperwidth][r]{CALT-TH-2025-001}}}}

\title{\Large\bf 
The Fermi function and the neutron's lifetime}

\author[1,2]{Peter Vander Griend}
\author[1]{Zehua Cao}
\author[1,2]{Richard~J.~Hill}
\author[3]{Ryan~Plestid}
\affil[1]{University of Kentucky, Department of Physics and Astronomy, Lexington, KY 40506 USA \vspace{1.2mm}}
\affil[2]{Fermilab, Theoretical Physics Department, Batavia, IL 60510 USA
\vspace{1.2mm}}
\affil[3]{Walter Burke Institute for Theoretical Physics, 
California Institute of Technology, Pasadena, CA 91125 USA\vspace{1.2mm}}

\date{\today}

\maketitle

\begin{abstract}
  \vspace{0.2cm}
  \noindent
  The traditional Fermi function ansatz for nuclear beta decay describes enhanced perturbative effects in the limit of large nuclear charge $Z$ and/or small electron velocity $\beta$.   We define and compute the quantum field theory object that replaces this ansatz for neutron beta decay, where neither of these limits hold. 
  We present a new factorization formula that applies in the limit of small electron mass, analyze the components of this formula through two loop order, and resum perturbative corrections that are enhanced by large logarithms. 
  We apply our results to the neutron lifetime, supplying the first two-loop input to the long-distance corrections. 
  Our result can be summarized as 
\begin{equation*}
    \tau_n \times  |V_{ud}|^2\big[1+3\lambda^2\big]\big[1+\Delta_R\big]
    = 
    \frac{5263.284(17)\,{\rm s}}
    {1 + 27.04(7)\times 10^{-3} }~,
\end{equation*}
with $|V_{ud}|$ the up-down quark mixing parameter, $\tau_n$ the neutron's lifetime, $\lambda$ the ratio of axial to vector charge, and $\Delta_R$ the short-distance matching correction. 
We find a shift in the long-distance radiative corrections compared to previous work, and discuss implications for extractions of $|V_{ud}|$ and tests of the Standard Model. 
\end{abstract}
\vfill

\section{Introduction}

The neutron's lifetime $\tau_n$ is an important precision observable within the Standard Model~\cite{Bopp:1986rt,Ando:2004rk,Darius:2017arh,Seng:2018yzq,Seng:2018qru,Fry:2018kvq,Czarnecki:2019mwq,Hayen:2020cxh,Seng:2020wjq,Gorchtein:2021fce,UCNt:2021pcg,Shiells:2020fqp}, offering a theoretically clean determination of the CKM matrix element $|V_{ud}|$ and probing physics beyond the Standard Model~\cite{Cirigliano:2013xha,Glick-Magid:2016rsv,Gonzalez-Alonso:2018omy,Glick-Magid:2021uwb,Falkowski:2021vdg,Brodeur:2023eul,Crivellin:2020ebi,Coutinho:2019aiy,Crivellin:2021njn,Crivellin:2020lzu,Cirigliano:2022yyo}. 
Unambiguous conclusions require control over radiative corrections~\cite{Sirlin:1967zza,Jaus:1972hua,Wilkinson:1982hu,Sirlin:1986cc,Jaus:1986te,Cirigliano:2022hob,Hill:2023acw,Hill:2023bfh,Borah:2024ghn,Seng:2024ker,Cirigliano:2024nfi}.  
It is well known that the neutron decay rate receives a large, $\sim 7\%$, first order QED radiative correction~\cite{Czarnecki:2004cw,Czarnecki:2019mwq}, two orders of magnitude larger than the naive expectation of $\alpha/(2\pi) \approx 10^{-3}$.  While a portion of this correction arises from electroweak logarithms and can be resummed by standard means,
the largest contributions arise
from the low-energy matrix element.

These numerically large 
long-distance
contributions have historically been estimated 
using a Fermi function ansatz;   
after integrating over phase space, the 
estimated corrections to the rate behave as~\cite{Fermi:1934hr,Wilkinson:1982hu,Czarnecki:2004cw}
\begin{align}\label{eq:FZW}
 1 + 4.6 \,\alpha + 16 \,\alpha^2 + 35 \, \alpha^3 + \dots \,.  
\end{align}
As we discuss below, the Fermi function 
does not give a controlled approximation 
to the complete decay amplitude beyond first order in $\alpha$. Nevertheless, the ansatz in Eq.~(\ref{eq:FZW}) predicts a permille-level contribution from the second-order term ($16\alpha^2$), which is larger than the precision goals for $|V_{ud}|$
determinations from neutron beta decay.  
What object should replace the ansatz in Eq.~(\ref{eq:FZW}) and its associated second-order correction?   

The Fermi function for beta decay~\cite{Fermi:1934hr} describes a class of 
enhanced radiative corrections arising from electron or positron propagation 
in a nuclear Coulomb field. 
The corrections are parametrically enhanced at large $Z$
and/or at small electron velocity $\beta$.  
However, neither limit holds for neutron beta decay: $Z$ is equal to 0 or 1 for the neutron or proton, and the region of small electron velocity is kinematically suppressed. 
In detail, the electron velocity spectrum is given by (for simplicity in this illustration we compute at tree level and in the heavy nucleon limit)
\begin{align} \label{eq:phase}
    {d\Gamma(n \to p e\nu)\over d\beta} \propto 
    {\beta^2 \over (1-\beta^2)^\frac52} \left[ {\Delta\over \me} - {1\over \sqrt{1-\beta^2}} \right]^2 \,,
\end{align}
where $m$ is the electron's mass, $\Delta = m_n-m_p$ is the difference between the neutron mass $m_n$ and the proton mass $m_p$, and 
the allowed range is $0 \le \beta \le \sqrt{1-\me^2/\Delta^2}$. 
The spectrum is strongly suppressed at small $\beta$: 
the mean velocity is $\langle \beta \rangle \approx 0.73$,
and less than 0.1\% of the total decay 
rate involves electron velocity $\beta <0.1$ (less than 10\% involves $\beta < 0.5$). 

In Refs.~\cite{Hill:2023acw,Hill:2023bfh}, two of us showed how the 
traditional Fermi function for nuclear beta decay may be identified as the leading-in-$Z$ contribution to a well-defined quantum field theory object (the hard contribution in the factorization formula for the process).  
Here, we show how the large first-order correction 
appearing in this quantum field theory object, and in 
Eq.~(\ref{eq:FZW}), arises from large logarithms
$|\log[(-E-\iu0)/E]| = \pi$, where $E$ is the electron energy. 
We show how these large logarithms are resummed using renormalization group (RG) methods.  
Moreover, 
in the limit of small electron mass 
(recall $\me^2/\Delta^2\approx 0.16$), we show that the Fermi function enhancement for neutron beta decay is governed by 
the universal cusp anomalous dimension for QED scattering amplitudes~\cite{Korchemsky:1987wg,Korchemskaya:1992je}.

In what follows, 
we construct a systematic description to replace Eq.~(\ref{eq:FZW}), 
identifying the object that replaces the traditional Fermi function ansatz for neutron beta decay as a component of a quantum field theory factorization formula. 
We decompose this object in the limit of small electron mass and compute the associated hard and jet functions through two loop order. 
Using these results, we present a new analysis of the long-distance radiative corrections to neutron beta decay, apply our formalism to obtain improved predictions for $\tau_n$ and $|V_{ud}|$, and comment on application of our results to nuclear beta decay.  

\section{Renormalization analysis}

In what follows, we will write the matrix element for neutron decay as a Dirac structure that acts between the electron and neutrino spinors. 
For an electron with momentum $p$ and a neutrino with momentum $k$, this object appears in the leptonic part of the amplitude as  $\bar{u}(p) \mathcal{M} \gamma_\mu (1-\gamma_5) v(k)$, and is normalized to unity at tree level,
$\mathcal{M}=1+O(\alpha)$. 
The matrix element including virtual photon corrections can be decomposed in terms of soft and hard virtual photon contributions i.e., $\mathcal{M}\sim \mathcal{M}_S\times \mathcal{M}_H$~\cite{Hill:2023acw}.
The product $\mathcal{M}_S\times \mathcal{M}_H$ thus encodes the long-distance contributions associated with physics below the scale of the nucleon masses; short-distance contributions are encoded in the Wilson coefficient obtained by matching onto the UV theory and are proportional to the weak vector and axial vector couplings $g_{V,A}$. 
The matching onto the UV theory is performed at the scale $\mu_{\rm UV}$.
Both the ratio of the vector and axial vector couplings and the product of the Wilson coefficient and the long-distance matrix element are independent of $\mu_{\rm UV}$.
Real photon contributions are discussed after Eq.~(\ref{eq:factorizationrate}).
The soft factor is known to all orders in perturbation theory \cite{Yennie:1961ad,Weinberg:1965nx}.  The hard matrix element can be computed order-by-order in perturbation theory.

Let us decompose the hard amplitude $\mathcal{M}_H$ in the static limit (defined as $\Delta/m_p\rightarrow 0$ with $\Delta$ held fixed). In this limit, we can label the proton and neutron with a conserved four-velocity $v_\mu$. With $v_\mu$ as an available reference vector, the amplitude can be written as
\begin{align}
    {\cal M}_H(w,\mu^2) &= {\cal A}_H(w,\mu^2) + {1\over w} \slash{v} {\cal B}_H (w,\mu^2) \,,
\end{align}
where $p^\mu$ is the electron four-momentum, $v^\mu=(1,0,0,0)$ in the neutron rest frame, 
$w=v\cdot p/m$,   and $\mu$ is the renormalization scale. Amplitudes computed in the static limit have certain simple properties. For instance taking $v\rightarrow -v$ (implying $w\rightarrow -w$) is equivalent to crossing (this follows immediately from the Feynman rules of the heavy-particle effective theory). Writing
\begin{align}\label{eq:refMplusminus}
    {\cal M}_H(w) = {\cal M}_H(-w) + \left[ {\cal M}_H(w)-{\cal M}_H(-w) \right] \,,
\end{align}
we recognize
the first object, ${\cal M}_H(-w)$, as the amplitude for the spacelike process where a heavy particle of charge $-1$ converts to an electron.\!\footnote{An example is the anti-particle analog of inverse beta decay i.e., $\nu_e \bar{p} \rightarrow e^- \bar{n}$.}  At one loop order, explicit calculation yields 
\begin{align}
    {\cal A}_H(-w) &= 1 + {\alpha\over 2\pi}
    \bigg[ \frac34 \log{\mu_{\rm UV}^2\over\me^2} + \log\frac{\mu^2}{\me^2} \left( wj(w)-1\right)
    + wj(w) - wJ(w) 
    \bigg] \,, \nl
    {\cal B}_H(-w) &= {\alpha \over 2\pi}\bigg[ -w j(w)\bigg] \,,
\end{align}
where (for $w>1$) the functions $j(w)$ 
and $J(w)$ are defined by 
\begin{align}
    wj(w) =& [w/\sqrt{w^2-1}]\log(w + \sqrt{w^2-1}),\nl
    wJ(w)=&[w/\sqrt{w^2-1}]\big[ {\rm Li}_2\left(1-(w-\sqrt{w^2-1})^2\right) + \log^2(w + \sqrt{w^2-1}) \big].
\end{align}

Again using properties of heavy-particle amplitudes, we observe that ${\cal M}_H(w)$ can be computed as the sum of the first object ${\cal M}_H(-w)$ plus all possible insertions of a $Z=+1$ background field~\cite{Plestid:2024eib,Borah:2024ghn} 
\begin{align}\label{eq:MHdiagrams}
    \vspace{6pt}
{\cal M}_H(w) &\equiv 
\hspace{-10mm}
\parbox{35mm}{
\begin{fmfgraph*}(100,55)
  \fmfleftn{l}{3}
  \fmfrightn{r}{3}
  \fmfbottomn{b}{5}
  \fmf{phantom}{l2,v,r2}
  \fmffreeze
  \fmf{plain}{r3,x,v}
  \fmf{double}{r1,y,v}
  \fmffreeze
\fmfv{decor.shape=circle,decor.filled=shaded,decor.size=5mm}{v}
\end{fmfgraph*}
}
\quad 
=
\quad
\left.\left( \parbox{37mm}{
\begin{fmfgraph*}(100,55)
  \fmfleftn{l}{3}
  \fmfrightn{r}{3}
  \fmfbottomn{b}{5}
  \fmf{phantom}{l2,v,r2}
  \fmffreeze
  \fmf{plain}{r3,x,v}
  \fmf{double}{l2,y,v}
  \fmffreeze
\fmfv{decor.shape=circle,decor.filled=shaded,decor.size=5mm}{v}
\end{fmfgraph*} 
}\right)\right|_{v\to -v}
\nl
&=\quad
\parbox{30mm}{
\begin{fmfgraph*}(100,55)
  \fmfleftn{l}{3}
  \fmfrightn{r}{3}
  \fmfbottomn{b}{5}
  \fmf{phantom}{l2,v,r2}
  \fmffreeze
  \fmf{plain}{r3,x,v}
  \fmf{double}{l2,y,v}
  \fmffreeze
\fmfv{decor.shape=circle,decor.filled=shaded,decor.size=5mm}{v}
\end{fmfgraph*} 
}
\quad
+
\quad
\parbox{30mm}{
\begin{fmfgraph*}(100,50)
  \fmfleftn{l}{3}
  \fmfrightn{r}{3}
  \fmfbottomn{b}{5}
  \fmf{phantom}{l2,v,r2}
  \fmffreeze
  \fmf{plain}{r3,x,v}
  \fmf{double}{l2,v}
  \fmffreeze
  \fmf{photon}{x,b4}
  \fmfv{decor.shape=cross}{b4}
    \fmfv{decor.shape=circle,decor.filled=full,decor.size=2mm}{v}
\end{fmfgraph*}
}
\quad + \quad 
\left(\rule{0cm}{1.4cm}\right.
\parbox{35mm}{
\begin{fmfgraph*}(100,50)
  \fmfleftn{l}{3}
  \fmfrightn{r}{3}
  \fmfbottomn{b}{7}
  \fmf{phantom}{l2,v,r2}
  \fmf{phantom}{l1,b5,b6,r1}
  \fmffreeze
  \fmf{plain}{r3,x,y,v}
  \fmf{double}{l2,v}
  \fmffreeze
  \fmf{photon}{x,b6}
  \fmf{photon}{y,b5}
  \fmfv{decor.shape=cross}{b5}
  \fmfv{decor.shape=cross}{b6}
      \fmfv{decor.shape=circle,decor.filled=full,decor.size=2mm}{v}
\end{fmfgraph*}
}  
 \nonumber \\[5mm]
& 
\quad + \quad
\parbox{30mm}{
\begin{fmfgraph*}(100,50)
  \fmfleftn{l}{3}
  \fmfrightn{r}{3}
  \fmfbottomn{b}{7}
  \fmf{phantom}{l2,v,r2}
  \fmf{phantom}{l1,b5,b6,r1}
  \fmffreeze
  \fmf{plain}{r3,x,y,v}
  \fmf{double}{l2,z,v}
  \fmffreeze
  \fmf{photon}{x,b6}
  \fmf{photon,right}{y,z}
  \fmfv{decor.shape=cross}{b6}
      \fmfv{decor.shape=circle,decor.filled=full,decor.size=2mm}{v}
\end{fmfgraph*}
}
\quad + \quad 
\parbox{30mm}{
\begin{fmfgraph*}(100,50)
  \fmfleftn{l}{3}
  \fmfrightn{r}{3}
  \fmfbottomn{b}{7}
  \fmf{phantom}{l2,v,r2}
  \fmf{phantom}{l1,b5,b6,r1}
  \fmffreeze
  \fmf{plain}{r3,x,y,v}
  \fmf{double}{l2,z,v}
  \fmffreeze
  \fmf{photon,right}{x,z}
  \fmf{photon}{y,b5}
  \fmfv{decor.shape=cross}{b5}
      \fmfv{decor.shape=circle,decor.filled=full,decor.size=2mm}{v}
\end{fmfgraph*}
}
\quad + \quad 
\parbox{35mm}{
\begin{fmfgraph*}(100,50)
  \fmfleftn{l}{3}
  \fmfrightn{r}{3}
  \fmfbottomn{b}{5}
  \fmf{phantom}{l2,v1,r2}
  \fmffreeze
  \fmf{plain}{r3,x,v1}
  \fmf{double}{l2,v1}
  \fmffreeze
  \fmf{photon}{x,a1}
  \fmf{photon}{a2,b4}
  \fmf{fermion,left,tension=0.5}{a1,a2,a1}
  \fmfv{decor.shape=cross}{b4}
      \fmfv{decor.shape=circle,decor.filled=full,decor.size=2mm}{v1}
\end{fmfgraph*}
}
\\[6pt] \nonumber
& \quad + \quad
\parbox{35mm}{
\begin{fmfgraph*}(100,50)
\fmfstraight
  \fmfleftn{l}{3}
  \fmfrightn{r}{3}
  \fmfbottomn{b}{5}
  \fmf{phantom}{l2,v1,r2}
  \fmffreeze
  \fmf{plain}{r3,y,z,w,v1}
  \fmf{double}{l2,v1}
  \fmffreeze
  \fmf{photon}{z,b4}
  \fmf{photon,right}{y,w}
  \fmfv{decor.shape=cross}{b4}
      \fmfv{decor.shape=circle,decor.filled=full,decor.size=2mm}{v1}
\end{fmfgraph*}
}
\quad + \quad
\parbox{35mm}{
\begin{fmfgraph*}(100,50)
    \fmfstraight
  \fmfleftn{l}{3}
  \fmfrightn{r}{3}
  \fmfbottomn{b}{8}
  \fmf{phantom}{l2,v1,r2}
  \fmffreeze
  \fmf{plain}{r3,y,z,w,v1}
  \fmf{double}{l2,v1}
  \fmffreeze
  \fmf{photon}{y,b7}
  \fmf{photon,right=1.5}{z,w}
  \fmfv{decor.shape=cross}{b7}
      \fmfv{decor.shape=circle,decor.filled=full,decor.size=2mm}{v1}
\end{fmfgraph*}
} \quad 
\left.\rule{0cm}{1.4cm}\right)
+ \dots \,.\\[6pt]
\nonumber
\end{align}
The second object in Eq.~(\ref{eq:refMplusminus}), ${\cal M}_H(w,\mu^2)-{\cal M}_H(-w,\mu^2)$, contains all diagrams with at least one background field insertion.  
At one loop order, 
\begin{align}\label{eq:1loopMH}
    {\cal A}_H(w)-{\cal A}_H(-w) &= {\alpha\over 2\pi} \bigg[ 
    {\iu\pi w\over\sqrt{w^2-1}}\left(  \log\left( -4\bm{p}^2-\iu0 \over \mu^2  \right) -1  \right) 
    \bigg] \,,
    \nl
    {\cal B}_H(w) - {\cal B}_H(-w) &= {\alpha\over 2\pi} \bigg[  {\iu\pi w\over\sqrt{w^2-1}} \bigg] \,, 
\end{align}
where we use 
\begin{align}
    \tilde{w}j(\tilde{w}) =& wj(w) -\iu\pi w/\sqrt{w^2-1},\nl
    \tilde{w} J(\tilde{w}) =& w J(w) - \iu\pi (w/\sqrt{w^2-1})\log(-4(w^2-1)-\iu0),
\end{align}
with $\tilde{w}=-w-\iu0$.

When $E/p = \beta^{-1}=w/\sqrt{w^2-1}$ and $E/m=w$ are order unity, there is no large ratio of physical scales, and naively no large logarithms in perturbation theory.  
However, even when there are no large ratios of scales, the difference of amplitudes ${\cal M}_H(w) - {\cal M}_H(-w)$ 
contains factors of $\log(-1-\iu0)\log[ (- 4\bm{p}^2-\iu0) / \mu^2] \sim -\pi^2$, 
{\it cf}. Eq.~(\ref{eq:1loopMH}). 
Such large logarithms are minimized by choosing $\mu^2= - 4\bm{p}^2-\iu0$ as opposed to, for example, $\mu^2=+4\bm{p}^2$~\cite{Ahrens:2008qu,Ahrens:2009cxz}.  Since the $\mu^2$ dependence of the amplitude is known to all orders, such enhancements can be resummed to all orders by RG methods, leading to the expression~\cite{Hill:2023acw} 
\begin{align}\label{eq:MHansatz}
{\cal M}_H(w,\mu^2) &= \exp\bigg[ {\pi\alpha\over 2\beta} + \iu \alpha \phi \bigg]  {\cal M}_H(w,-\mu^2-\iu0) \,, 
\end{align}
where 
\begin{align}
    \phi= \frac12[ wj(w)-1] =
\frac12\left( \frac{1}{2\beta}\log{1+\beta\over 1-\beta} -1   \right).
\end{align}
In the sections that follow, we show how the expression (\ref{eq:MHansatz}) emerges, to all orders in perturbation theory, in the small-$m$ limit. Furthermore, this same analysis relates the Fermi function to enhancements that stem from the RG evolution of the universal gauge theory cusp anomalous dimension.   
We compute ${\cal M}_H(w,-\mu^2-\iu0)$, including virtual contributions through two-loop order, and estimating residual corrections from real radiation and 
$\me^2/\Delta^2$
power corrections.  
We include relevant recoil corrections and the UV matching coefficient to arrive at updated predictions for $\tau_n$ and $|V_{ud}|$ from the neutron lifetime. 

 \section{Factorization and small mass expansion}

Using the expression (\ref{eq:MHansatz}) and the complete one-loop result for ${\cal M}_H(w,-\mu^2-\iu0)$, we find remaining corrections are of order $\alpha^2$, without logarithmic enhancements arising from the scale $\mu^2\sim -4\bm{p}^2$. 
In order to investigate the numerical convergence of perturbation theory for 
the object ${\cal M}_H(w,-\mu^2-\iu0)$ and to further clarify the physical meaning
of a ``Fermi function" for neutron beta decay, let us work to leading power in 
the small-$\me$ expansion, $\me^2/\Delta^2 \approx 0.156$, where $\Delta$
is the maximal electron energy in the static limit.    
In the small-$\me$ limit, the virtual corrections to neutron beta decay factorize~\cite{Hill:2016gdf,longpaper}
\begin{align}\label{eq:factorization}
{\cal M}_H(w) \approx A_H(w) \approx 
F_R(w,m) F_J(m) F_H(E) \,. 
\end{align}
Apart from the small correction from the ``remainder" function $F_R$, which converts between $n_{e}=1$ and $n_{e}=0$ dynamical electrons in the low-energy effective theory, the amplitude factorizes into a collinear or ``jet" function $F_J$ depending only on the mass scale $m$, and a ``hard" function $F_H$ depending only on
the energy scale $E$.

In terms of the well-behaved ${\cal A}_H(-w)$, enhancements
are contained in the ratio 
\begin{align}
\left| { {\cal A}_H(w) \over {\cal A}_H(-w)} \right| &\approx \left| F_H(E) \over F_H(-E) \right| \,, 
\end{align}
where we use that $F_R(w)/F_R(-w)$ is a pure phase~\cite{Hill:2016gdf},
{\it cf}. Eq.~(\ref{eq:two-loop-Fs}). 
Since $F_H$ (upon setting $\mu_{\rm UV}=\mu$) 
depends only on the dimensionless ratio $E/\mu$ and since the $\mu$ dependence is determined by renormalization, we can resum enhancements in this ratio~\cite{longpaper}. 
At the scale $\mu=\mu_*=2E$,
\begin{align}\label{eq:FHratio}
    \left| F_H(E, \mu_* ) \over F_H(-E, \mu_*)\right|^2 
    = \left| F_H(E, \mu_*) \over F_H(E, -\mu_*-\iu0) \right|^2
    = \exp\bigg[ -X_*^2 {\bar{\alpha}\over 4\pi} +  \frac{32}{9}n_{e} X_*^2  \left({\bar{\alpha}\over 4\pi} \right)^2 -\frac{8}{27} n_{e}^2 X_*^4  \left({\bar{\alpha}\over 4\pi} \right)^3
    + \dots
    \bigg] \,,
\end{align}
where $X_* = \log{\mu_*^2 \over (-\mu_*-\iu0)^2} = 2\pi \iu$.  Here $\overline{\alpha}$ is the $n_{e}=1$ flavor $\overline{\rm MS}$ coupling and is given in terms of $n_{e}=0$ (on-shell) $\alpha$ as
\begin{equation}\label{eq:alpha_msbar}
    \overline{\alpha} = \alpha \left( 1 -\frac{4n_{e}}{3} \frac{\alpha}{4\pi}  \log{\me^2\over \mu^2} + \dots  \right).
\end{equation}
We thus recover the exponential factor in Eq.~(\ref{eq:MHansatz}), 
together with a series of subleading logarithms. 

After isolating the factor (\ref{eq:FHratio}), we may use explicit two-loop results extracted~\cite{longpaper} from the literature~\cite{Broadhurst:1994se,Grozin:1998kf,Bekavac:2009zc,Beneke:2008ei,Bonciani:2008wf,Hill:2016gdf} to compute the remaining components of the factorization formula through two-loop order:
\begin{align}
    \label{eq:two-loop-Fs}
      F_R(-w,m,\mu) &= 1 + \left(\bar{\alpha} \over 4\pi \right)^2 \left(\log(2w)-1\right) n_{e} 
    \left( -\frac43 L_m^2 - \frac{40}{9}L_m - \frac{112}{27} \right) \,,
    \nl
     F_J(m,\mu) &= 1 + {\bar{\alpha}\over 4\pi}
    \left[ \frac12 L_m^2 -\frac12 L_m + 2 + {\pi^2\over 12} \right]
\nl
&\quad +  \left({\bar{\alpha}\over 4\pi}\right)^2\bigg[ 
    \frac18 L_m^4 
   -\frac16 \left( - \frac43 n_{e} + \frac32 \right) L_m^3 
    -\frac14\left( \frac{52}{9} n_{e}  - \frac92 - {\pi^2\over 6} \right) L_m^2 
    \nl
    &\quad 
    -\frac12 \left( n_{e}\left( -{154\over 27} - {8\pi^2\over 9}\right)
    + \frac72 - {23\pi^2\over 12} + 24\zeta_3
    \right)L_m
    \nl
    &\quad 
+ n_{e} \left( {4435\over 324} - \frac29 \zeta_3 -{41\pi^2\over 54}\right)
   + 2\pi^2 -4\pi^2\log{2} - {331\pi^4\over 1440} - 3\zeta_3 + {209\over 16}
    \bigg] \,,
    \nl
    F_H(-E,\mu) &= 1 +  {\bar{\alpha}\over 4\pi}\left[ -2 L_E^2  + 2 L_E - 2 - {5\pi^2 \over 12} \right]
    \nl
    &\quad 
    + \left( {\bar{\alpha}\over 4\pi}\right)^2
    \bigg\{
n_{e}\Bigg[ -\frac{16}{9}L_E^3 + \frac{64}{9}L_E^2 
+ \left(-\frac{304}{27}-\frac{16\pi^2}{9} \right)L_E
    + {656\over 81} + \frac29\zeta_3
    + {113\pi^2 \over 54}
    \bigg] \nl
    &\quad
    + 2L_E^4 - 4L_E^3 
    + \left(6 + {5\pi^2\over 6}\right)L_E^2
    + \left( 24\zeta_3- {11\pi^2\over 2}\right) L_E - 8 + {65\pi^2\over 6}
    - {167\pi^4\over 288} - 15 \zeta_3
 \Bigg\} \,,
\end{align}
 where $L_m \equiv \log(\me^2/ \mu^2)$ and $L_E=\log(2E/\mu)$.
The above components of the factorization formula can be expressed in terms of on-shell $\alpha$ using Eq.~(\ref{eq:alpha_msbar}).

\section{Radiative corrections in the static limit \label{sec:Static-Limit}}

We first present radiative corrections in the static limit, in terms of the tree-level decay rate.  These corrections are our main focus; in the following section, we combine these corrections with known 
recoil corrections and express the tree-level rate in terms of weak-interaction couplings to determine the neutron lifetime. 

The differential neutron beta decay rate (including final state photons) can be expressed as 
\begin{align}\label{eq:factorizationrate}
{\dd\Gamma\over \dd E} &= \left( \dd \Gamma\over \dd E\right)_{\! \rm tree} 
S(\deltaE,\mu^2) \, H(\deltaE, \mu^2) \,,
\end{align}
where $\deltaE$ is a soft-photon energy cutoff defined in the rest frame of the neutron. 
Dependence on $\deltaE$ cancels  between $S$ and $H$ order-by-order in $\alpha$ when all real and virtual photon effects are included.
When $\deltaE$ is assumed small, the soft function exponentiates
\begin{align}
    \log S(\deltaE) &=
    {\alpha\over 2\pi}\bigg[
     \log{2\deltaE\over \mu} \left( {2\over \beta}\log{1+\beta\over1-\beta}-4\right)
    - {1\over 2\beta} \log^2\left(1+\beta\over 1-\beta\right) 
    -{2\over\beta}{\rm Li}_2\left(2\beta\over 1+\beta\right) + {1\over\beta}\log{1+\beta\over 1-\beta} 
    +2
    \bigg] \,.
\end{align}
Employing Eq.~(\ref{eq:MHansatz}), we write
\begin{align}\label{eq:Hfull}
    H(\deltaE, \mu^2) &=  
     1 +  {\alpha\over 2\pi}{H}^{(1)} + \left(\alpha\over 2\pi\right)^2 {H}^{(2)} + \dots 
    \nl
    &= \exp\bigg[ {\pi\alpha\over \beta}\bigg] H(\deltaE, -\mu^2-\iu0) 
    \\
    &= \left(1 + {\pi\alpha\over \beta} + {\pi^2\alpha^2\over 2\beta^2} \right)
    \left[ 1 +  {\alpha\over 2\pi}\hat{H}^{(1)} + \left(\alpha\over 2\pi\right)^2 \hat{H}^{(2)} 
    \right] + \dots \,. \nonumber
\end{align}
The quantities $H^{(n)}$ and $\hat{H}^{(n)}$ are expansion coefficients for $H(\deltaE,\mu^2)$ and $H(\deltaE,-\mu^2-\iu0)$ respectively. 
We may further consider the expansion in electron mass
\begin{align}
    H^{(n)} &= \big[H^{(n)}\big]_0 + \left({\me^2\over \Delta^2}\right)\big[H^{(n)}\big]_1 + \left(\me^2\over \Delta^2\right)^2 \big[ H^{(n)}\big]_2 + \dots \,,
    \nl
     \hat{H}^{(n)} &= \big[\hat{H}^{(n)}\big]_0 + \left({\me^2\over \Delta^2}\right)\big[\hat{H}^{(n)}\big]_1 + \left(\me^2\over \Delta^2\right)^2 \big[\hat{H}^{(n)}\big]_2 + \dots \,.
\end{align}
For $H^{(1)}$ and $\hat{H}^{(1)}$, we use the well known exact result~\cite{Sirlin:1967zza}
\begin{align}
    H^{(1)} &= H^{(1)}_V + H^{(1)}_R \,,
\end{align}
with the virtual and real photon contributions
\begin{align}
       H_V^{(1)}&= 3\log{\mu_{\rm UV}\over m} + \log{\mu\over m}\left( {2\over \beta}\log{1+\beta\over 1-\beta} - 4\right) 
    + \beta \log{1+\beta\over 1-\beta} + {2\pi^2\over \beta} - {2\over\beta}{\rm Li}_2\left(2\beta\over 1+\beta\right) - 
\frac{1}{2\beta}\log^2\left(1+\beta\over 1-\beta\right) \,,
     \nl
     H_R^{(1)} &= \log{\deltaE \over \Delta - E} \left(4 - {2\over \beta}\log{1+\beta\over 1-\beta} \right) 
     + 
\frac{1}{\beta}
\log{1+\beta\over 1-\beta}\left[ {(\Delta-E)^2\over 12 E^2} 
     + {2(\Delta-E)\over 3E}- 3
     \right]
     - {4(\Delta-E)\over 3E} + 6 \,.
     \label{H1-exprs}
\end{align}
For virtual photon contributions to 
$[H^{(2)}_V]_0$ and $[\hat{H}^{(2)}_V]_0$  we
equate the small mass limit of
Eq.~(\ref{eq:Hfull}) 
with the expression obtained using Eq.~(\ref{eq:factorization}).

Table~\ref{tab:numerics} examines the convergence of perturbation theory for the hard function, comparing the direct expansion (i.e., $H^{(n)}$) with the expansion after extracting $\exp(\pi\alpha/\beta)$ (i.e., $\hat{H}^{(n)}$).  In both cases, the complete soft function is included to all orders.
We estimate the impact of omitted real radiation contributions at two-loop order by including the known $\deltaE$ dependence as 
$\log[\deltaE/\Lambda_{\gamma}]$.
The resulting hard function is independent of $\deltaE$, and finite terms are estimated by varying $\Lambda_{\gamma} = \Delta/2 \dots 2\Delta$. 
For comparison, Table \ref{tab:numerics} shows the analogous exercise at one-loop order.   
Neglected higher-order perturbative corrections are
estimated by renormalization scale variation: $\mu = m/2 \dots 2\Delta$.  
Finally, for an estimate of neglected power corrections 
$[H^{(2)}]_i$, $[\hat{H}^{(2)}]_i$, for 
$i\ge 1$, we first observe that these corrections contribute to the difference between left and right columns in the last row of the table: $29.31-29.04=0.27$.

For a more refined estimate of power corrections, we consider the inclusion of a gauge invariant subclass of power corrections represented by the first two-loop diagram on 
the right hand side of Eq.~(\ref{eq:MHdiagrams}).  This class of diagrams (photon exchange with an external field) contains the leading $1/\beta^2$ dependence of the hard function  
at $\beta\to 0$,
amounting to the following power correction
\begin{align}
    \label{eq:FF-power-correction}
    \hat{H}^{(2)}_V - \big[\hat{H}^{(2)}_V\big]_0 
    \approx - {2\pi^4 \over 3}{\me^2\over E^2-\me^2} 
    = - {2\pi^4 \over 3}\left( {\me^2\over E^2} + {\me^4\over E^4} + \dots  \right)
    \,,
\end{align}
and shifts the central value for the 
``With Resummation'' column of Table \ref{tab:numerics} as 
$29.31 \to 29.18$.
We assign a residual power correction uncertainty as $1/2$ of this shift.
The remaining two-loop diagrams in Eq.~(\ref{eq:MHdiagrams}) have been estimated to contribute at a numerically small ($10^{-5}$) level~\cite{Sirlin:1986cc}, after including the iteration of one-loop subdiagrams, which are automatically incorporated in our resummed analysis  by employing $\hat{H}$ in place of $H$.  

We define the radiative corrections in the static limit, $\delta_{R,{\rm static}}$,  
by $\Gamma_{\rm static} = (1+\delta_{R,{\rm static}})\times (\Gamma_{\rm static})_{\rm tree}$ 
and take as our final result (at a renormalization scale $\mu_{\rm UV} = \Delta$),
\begin{align}\label{eq:deltaRstatic}
    \delta_{R,{\rm static}}(\mu_{\rm UV}=\Delta)
    =& \, ( 29.18 \pm 0.07 \pm 0.01 \pm 0.02 ) \times 10^{-3}\,.
\end{align}
The uncertainties are, respectively, from neglected power corrections in $\me^2/\Delta^2$ at two-loop order, from neglected real radiation at two-loop order, and from perturbative corrections at three-loop order.
We now turn to the application of our result to the neutron's lifetime. 

\begin{table}[t]
\centering
\begin{tabular}{|l||*{2}{c|}}\hline\hline
& Without Resummation & With Resummation \\
\hline\hline
$1$ & 
$\phantom{3}0.3\phantom{0} \pm 3.5\phantom{0} \pm 2.1\phantom{0}$ & 
$34.5\phantom{00} \pm 3.6\phantom{00} \pm 2.2\phantom{0}$
\\
$1 + H^{(1)}_V$ & $32.6\phantom{0} \pm 0.1\phantom{0} \pm 2.2\phantom{0}$  &
$33.2\phantom{00} \pm 0.004 \pm 2.2\phantom{0}$ \\ 
$1 + H^{(1)}$ & $28.8\phantom{0} \pm 0.08 \pm 0.05$ &
$29.32\phantom{0} \pm 0.02\phantom{0} \pm 0.01$
\\
$1 + H^{(1)} + H^{(2)}_V$ &
$29.04 \pm  0.05 \pm 0.05$ &
$29.31\phantom{0}\pm 0.02\phantom{0} \pm 0.01$
\\[3pt]\hline
\end{tabular}
\hfill
\begin{tabular}{|l||*{1}{c|}}\hline\hline
Quantity & Value $[10^{-3}]$ \\
\hline\hline
$\Delta_R$ & $\phantom{-}45.37 \pm 0.27$ \\
$\delta_{R,{\rm static}}$ & $\phantom{-}29.18 \pm 0.07$ \\
$\delta_{\rm recoil}$ & $\phantom{4}-2.06\phantom{\pm0.08}\phantom{4}\hspace{4pt}$ \\
$\delta_{\rm rad.rec.}$ & $\phantom{4}-0.08\phantom{\pm0.08}\phantom{4}\hspace{4pt}$ 
\\[3pt]\hline
\end{tabular}
\caption{
\textbf{(Left)} Long-distance radiative correction ($\delta_{R, {\rm static}}$) to the neutron decay rate, computing the hard function at different orders in perturbation theory, 
in units of $10^{-3}$.
The first column of numbers shows direct expansion of the hard function, and the second column shows the expansion after extracting 
$\exp(\pi\alpha/\beta)$ as in Eq.~(\ref{eq:Hfull}).  Central values are evaluated at $\mu^2=\me \Delta$, $\Lambda_\gamma = \Delta$, and $\mu_{\rm UV}=\Delta$ while the errors denote scale variation $\mu = m/2..2\Delta$, and $\Lambda_{\gamma} = \Delta/2 .. 2\Delta$ as discussed beneath Eq.~(\ref{H1-exprs}). 
\textbf{(Right)} Summary of radiative and recoil corrections to neutron decay rate.  
Electroweak and QED corrections $\Delta_R$ and $\delta_{R}$ are evaluated at renormalization scale $\mu_{\rm UV} = \Delta$, in the $\overline{\rm MS}$ scheme with $n_{e}=1$ dynamical electron. The power correction given in Eq.~(\ref{eq:FF-power-correction}) accounts for the shift $29.31\rightarrow 29.18$ in going from the left table to the right table for $\delta_{R,{\rm static}}$.\label{tab:numerics} 
} 
\end{table}

\section{Neutron lifetime}

The neutron lifetime is given by 
\begin{equation}\label{eq:Gamman}
    \Gamma_n =  \frac{ G_F^2|V_{ud}|^2\Delta^5}{ 2\pi^3} 
    f_{\rm static} 
    (1+3\lambda^2) 
    \bigg[1+\Delta_R(\mu_{\rm UV})\bigg] 
\bigg[1+\delta_{R,\rm static}(\mu_{\rm UV}) + \delta_{\rm recoil}+ \delta_{\rm rad.rec.} \bigg]\,,
\end{equation}
where $\lambda=g_A/g_V$ is the ratio of axial to vector weak-couplings of the nucleon, and the phase space factor in the static limit is given by
\begin{equation}\label{eq:fstatic}
    \begin{split}
    f_{\rm static}&= \int_{\me/\Delta}^1 \dd y~ y~(1-y)^2
    \sqrt{y^2-\qty(\tfrac{\me}{\Delta})^2}  \simeq 
    0.0157528 \,.
    \end{split}
\end{equation}
The short-distance radiative correction 
\begin{equation}
    \Delta_R(\mu_{\rm UV})\equiv g_V^2(\mu_{\rm UV})-1,
\end{equation} 
encodes short-distance electroweak and hadronic physics 
above the scale $\mu_{\rm UV}$~\cite{Czarnecki:2004cw,Seng:2018qru,Seng:2018yzq,Czarnecki:2019mwq, Cirigliano:2023fnz}; we do not include an estimate for isospin breaking in $\Delta_R$ since its numerical value ($\sim -4\times 10^{-5}$ \cite{Kaiser:2001yc}) is roughly six times smaller than the current error estimate on $\Delta_R$. The term $\delta_{R,{\rm static}}$ encodes the long-distance radiative corrections (as given above), 
and $\delta_{\rm recoil}$ and $\delta_{\rm rad.rec.}$ are recoil and radiative recoil corrections.  The recoil corrections are computed as described in Refs.~\cite{Ando:2004rk,Wilkinson:1982hu}, and we include the effect of the induced pseudoscalar form factor (i.e., one-pion exchange) \cite{Wilkinson:1982hu}. The radiative-recoil correction includes 
the dominant
interference between recoil terms and the first-order 
$\pi \alpha/\beta$ correction, and the shift between the electron velocity in the proton versus neutron rest frame~\cite{Wilkinson:1982hu}. 
A summary of recoil and radiative recoil corrections is given in the Supplemental Material.

In terms of $|V_{ud}|$, $\lambda$, and $\Delta_R(\mu_{\rm UV}=\Delta)$
the neutron lifetime is thus given by (restoring $\hbar$ for SI units and using inputs for $m_n$, $m_p$, $G_F$ from the Particle Data Group (2024)~\cite{ParticleDataGroup:2024cfk})\footnote{The normalization factor $\Delta^5 f_{\rm static}$ 
is defined in the static limit, {\it cf.} Eq.~(\ref{eq:fstatic}), 
and differs from the quantity $m_e^5 f_0$ used in  
Ref.~\cite{Cirigliano:2023fnz}, Eq.~(4). 
We combine this difference with other recoil corrections in our $\delta_{\rm recoil}$.  The total effect of recoil corrections is the same in our accounting as in Ref.~\cite{Cirigliano:2023fnz} up to subleading corrections, {\it cf.} the discussion after Eq.~(\ref{eq:CDMT}) below.}
\begin{equation}\label{eq:taun}
    \tau_n \times  |V_{ud}|^2(1+3\lambda^2)\bigg[1+\Delta_R(\mu_{\rm UV}=\Delta)\bigg] 
    \bigg[1 + 27.04(7)\times 10^{-3} \bigg]
    = \frac{2\pi^3 \hbar}{G_F^2 \Delta^5 f_{\rm static}} = 
    5263.284(17)
    \,{\rm s}~.
\end{equation}
As an illustrative example, let us take the lifetime of the neutron from the most recent UCN$\tau$ average, $\tau_n=877.82(30)\,{\rm s}$ \cite{Musedinovic:2024gms} and the measurement of $\lambda$ from the PERKEO-III experiment~\cite{Markisch:2018ndu}, $\lambda=-1.27641(56)$.  
Using $\Delta_R(\mu_{\rm UV}=\Delta) = 45.37(27)\times 10^{-3}$ \cite{Cirigliano:2023fnz},\!\footnote{This value for $\Delta_R$ is taken from Ref.~\cite{Cirigliano:2023fnz} (see also Table 2 of Ref.~\cite{Cirigliano:2022yyo} and Refs.~\cite{Seng:2018yzq,Seng:2018qru,Czarnecki:2019mwq,Seng:2020wjq,Hayen:2020cxh,Shiells:2020fqp}). We have converted between the renormalization scheme of Ref.~\cite{Cirigliano:2023fnz} and conventional $\overline{\rm MS}$ at renormalization scale $\mu=\Delta$.} we obtain
\begin{equation}\label{eq:Vud}
    \begin{split}
    |V_{ud}| &= 0.97393(17)_\tau (35)_\lambda (13)_{\Delta_R} (3)_{\delta_R} \\
    &= 0.97393 (41) \,,
    \end{split}
\end{equation}
where in the final line, errors have been added in quadrature.
Using average values from Ref.~\cite{ParticleDataGroup:2024cfk} 
for $\tau_n$ 
($878.4(5){\rm s}$ excluding beam measurements or $878.6(6){\rm s}$ including beam measurements) in place of the most precise measurement 
($\tau_n=877.82(30){\rm s}$~\cite{Musedinovic:2024gms})
yields a similar result in Eq.~(\ref{eq:Vud}) ($\sim 1\sigma$ downward shift in $|V_{ud}|$ and similar total error). 
Using the average from Ref.~\cite{ParticleDataGroup:2024cfk} 
for $\lambda$ ($-1.2754(13)$) in place of the most precise measurement ($\lambda = -1.27641(56)$~\cite{Markisch:2018ndu})
yields a consistent central value, and approximately 
two times larger total error.  
An in-beam measurement of $\tau_n$~\cite{Yue:2013qrc} is $\sim 4\sigma$ discrepant with the ultracold neutrons (UCN) measurements, which dominate the average.\!\footnote{
The in-beam-measurement of $\tau_n$ also yields a 
value for $|V_{ud}|$ that is  discrepant with determinations 
from superallowed beta decays~\cite{Hardy:2020qwl}.}
For a discussion of the discrepancy between in-beam and UCN measurements of $\tau_n$, see Refs.~\cite{Czarnecki:2018okw,Fornal:2018eol}. 
We have computed radiative corrections to the decay rate for the process
$n\to p e\bar{\nu}(\gamma)$.  This rate determines the neutron lifetime in the Standard Model, but should be interpreted as a partial rate if neutron decay modes beyond the Standard Model are present.

\section{Discussion}

Our new result, Eq.~(\ref{eq:deltaRstatic}), modifies the long-distance radiative correction to neutron beta decay.  Compared to previous work~\cite{Czarnecki:2019mwq,Cirigliano:2023fnz}, the largest effect corresponds to the replacement of the Fermi function ansatz, 
\begin{align} 
\label{eq:FNR-ansatz}
    F_{\rm NR} = {{(2\pi\alpha/\beta)}\over 1 - \exp(-2\pi\alpha/\beta)}
    \xrightarrow[\me\to 0]{}
1 + {\pi\alpha} + {\pi^2\alpha^2\over 3} + \dots \,,
\end{align}
with the resummation (\ref{eq:MHansatz}), 
\begin{align}
\label{eq:resum-factor}
    \left| {\cal M}_H(\mu^2) \over {\cal M}_H(w,-\mu^2-\iu0) \right|^2 
    &= \exp\bigg[ {\pi\alpha\over \beta}\bigg]  
     \xrightarrow[\me\to 0]{}
    1 + {\pi\alpha} + {\pi^2\alpha^2\over 2} + \dots \,.
\end{align}
We have presented the first complete analysis of the two-loop virtual corrections in the limit of small $\me^2/\Delta^2$, included 
leading contributions and 
uncertainties associated with power corrections 
and real radiation
at two loop order, and included all relevant recoil and radiative recoil corrections. Keeping full mass dependence at tree-level and one-loop, and upon including these corrections, the decay spectrum has the correct $\beta\rightarrow 0$ and $(\me^2/E^2)\rightarrow 0$ limits through two-loop order. 

Let us return to the 
ansatz (\ref{eq:FZW}) 
(evaluated at $R=1.0\,{\rm fm}$~\cite{Wilkinson:1982hu}), 
\begin{align}
   & 1 + \left\langle {\pi\over \beta}\right\rangle \alpha
    + \left\langle {\pi^2\over 3\beta^2} + {11\over 4}
    - \log(2p R\exp{\gamma_E}) 
    \right\rangle \alpha^2  + \dots 
    \nl
    &\quad
    \approx 1 + 4.6\,\alpha + \big( 8.3 + 2.8 + 4.6 \big)\alpha^2 + \dots \,,
\end{align}
where angle brackets denote  averaging over phase space.
We may now clarify the contributions to the  coefficient of $\alpha^2$.
First, we note that the contribution involving $\log(2p/\mu_R)$  
($\mu_R=\exp(-\gamma_E)R^{-1}$) represents the renormalization between scales $p \sim \Delta$ and $\mu_R \sim m_p$ arising from the $Z^2$ part of the anomalous dimension: in the notation of Ref.~\cite{Borah:2024ghn}, 
$\gamma_1 = \gamma_1^{(0)} Z^2 +\gamma_1^{(1)} Z + \gamma_1^{(2)}$, 
the coefficient of $\log(R^{-1})$ corresponds to 
$32\pi^2 \gamma_1^{(0)}$.  For neutron beta decay however, only the local heavy-light current contribution $\gamma_1^{(2)}$ survives 
($Z=0$ for the neutron), so that this contribution is spurious; the complete renormalization group running is known to high precision and included in our analysis ({\it cf}. Ref.~\cite{Cirigliano:2023fnz} for a related discussion).  
The remaining two-loop contributions, $\pi^2/(3\beta^2) + 11/4$,
are replaced by our systematic analysis of the low-energy matrix element.

Our results provide the first $O(\alpha^2)$ input (beyond the Fermi-function ansatz) to neutron beta decay.  
Let us compare to existing results in the literature. 
For example, in Eqs.~(107) and~(113) of Ref.~\cite{Cirigliano:2023fnz}, the long-distance contribution in Eq.~(\ref{eq:Gamman}) is given by 
\begin{align}\label{eq:CDMT}
    \bigg[1+\delta_{R,\rm static}(\mu_{\rm UV}) + \delta_{\rm recoil}+ \delta_{\rm rad.rec.} \bigg]^{\rm CDMT}
    &= 1 + 26.934(50)\times 10^{-3} \,.
\end{align}
which represents a shift of $1.1\times 10^{-4}$ ($2.1\sigma$) compared to our central value for the long-distance correction, $27.04 \times 10^{-3}$ in Eq.~(\ref{eq:taun}).
We note that the majority of this difference originates from the replacement of the nonrelativistic Fermi function (\ref{eq:FNR-ansatz}) by the resummation factor (\ref{eq:resum-factor}); smaller differences arise from other two-loop contributions included in the present analysis, and from subleading radiative recoil terms of similar size included in the factorization ansatz of Ref.~\cite{Cirigliano:2023fnz}.
In Eq.~(15) and line 5 of Table~1 of Ref.~\cite{Czarnecki:2019mwq}, 
\begin{align}
    \bigg[1+\delta_{R,\rm static}(\mu_{\rm UV}) + \delta_{\rm recoil}+ \delta_{\rm rad.rec.} \bigg]^{\rm CMS}
    &= 1 + 26.33(33) \times 10^{-3} \,,
\end{align}
which represents a shift of $7.1\times 10^{-4}$ ($2.2\sigma$) 
compared to our result given in Eq.~(\ref{eq:taun}).

In order to compare with alternative determinations of the short-distance corrections from the literature~\cite{Czarnecki:2019mwq}
we use the conversion formula, 
\begin{equation}
    1 + \Delta_R(\mu_{UV}=\Delta) =
    R(\Delta,m_p) 
    \bigg( 1 + \Delta_R^V - 
    {11\alpha \over 8\pi}
    \bigg)\,,
\end{equation}
where $\Delta_R^V$ is defined in the convention of Ref.~\cite{Czarnecki:2019mwq}, 
$R(\Delta,m_p)=1.0234650$ relates renormalization scales 
$\mu=\Delta$ and $\mu=m_p$,
and the factor $-11\alpha/(8\pi)$ converts the Sirlin convention 
for one-loop corrections
to $\overline{\rm MS}$~\cite{Hill:2023acw}. 
Taking $\Delta_R^V=0.02426(32)$ from Ref.~\cite{Czarnecki:2019mwq}
yields $\Delta_R(\mu_{\rm UV}=\Delta)= 45.03(33) \times 10^{-3}$;
with the same inputs for $\tau_n$ and $\lambda$, this translates 
to 
$|V_{ud}|=0.97409(17)_\tau (35)_\lambda (15)_{\Delta_R} (3)_{\delta_R}$.

We have 
provided the first two-loop input to the long-distance radiative corrections to neutron beta decay.
The result, Eq.~\eqref{eq:taun}, 
sets a target for uncertainty reductions in the experimental and short-distance inputs.
Existing determinations of the neutron's lifetime $\tau_n$ and axial-vector charge ratio $\lambda$ have already reached a level where errors on $|V_{ud}|$ are competitive with, albeit still larger than, superallowed beta decays. In order to surpass superallowed data, measurements of $\tau_n$ must reduce their errors by a factor of roughly two, while a reduction in the uncertainty on $\lambda$ requires a reduction in uncertainty by at least a factor of three \cite{Alarcon:2023gfu}. Proposals at the European Spallation Source suggest that the necessary reduction in the error on $\lambda$ is achievable \cite{Abele:2022iml}, while the UCN$\tau$+ upgrade will offer the necessary reduction in uncertainty on $\tau_n$ \cite{Alarcon:2023gfu}. In fact, a recent update from UCN$\tau$ \cite{Musedinovic:2024gms} has already achieved a small reduction in the uncertainty on the neutron lifetime as compared to their 2021 dataset average \cite{UCNt:2021pcg}. In conjunction with these experimental advances, progress towards a lattice-QCD determination of $\Delta_R$ has recently been undertaken \cite{Seng:2019plg,Seng:2020wjq,Feng:2020zdc,Ma:2023kfr}. 

We note that our two-loop analysis can also be applied to other observables (such as asymmetry coefficients) for future precision determinations of $\lambda$; currently radiative corrections are not a dominant source of uncertainty for extractions of $\lambda$ \cite{Markisch:2018ndu}. Finally, let us note that Eqs.~(\ref{eq:MHdiagrams}) and (\ref{eq:two-loop-Fs}) can be used to a determine the low-energy $Z\alpha^2$ matrix element in the recently developed effective field theory framework \cite{Hill:2023acw,Hill:2023bfh,Borah:2024ghn,Plestid:2024eib,Cirigliano:2024nfi,Cirigliano:2024msg} that is currently an important source of uncertainty in the analysis of superallowed beta decays. This would supplant previous estimates computed in the independent particle model \cite{Jaus:1972hua,Sirlin:1986cc,Sirlin:1986hpu,Jaus:1986te,Cirigliano:2024msg}.

\end{fmffile}

\section*{Acknowledgments}
We thank Vincenzo Cirigliano and Bradley Filippone for helpful discussions. We thank Susan Gardner, Martin Hoferichter, and Andreas Kronfeld for comments on the manuscript. 
 PVG acknowledges support from the Visiting Scholars Award Program of the Universities Research Association.
 RP thanks the Institute for Nuclear Theory at the University of Washington for its kind hospitality and stimulating research environment during program INT 23-1b.
 This research was supported in part by the INT's U.S. Department of Energy grant No. DE-FG02-00ER41132. Part of the research of RP was performed at the Kavli Institute for Theoretical Physics which is supported by the National Science Foundation under Grant No. NSF PHY-1748958.
 RP is supported by the Neutrino Theory Network under Award Number DEAC02-07CH11359, the U.S. Department of Energy, Office of Science, Office of High Energy Physics under Award Number DE-SC0011632, and by the Walter Burke Institute for Theoretical Physics.
This work was supported by the U.S. Department of Energy, Office of Science, Office of High Energy Physics, under Award DE-SC0019095.
This work was produced by Fermi Forward Discovery Group, LLC under Contract No. 89243024CSC000002 with the U.S. Department of Energy, Office of Science, Office of High Energy Physics.
RJH gratefully acknowledges support from the HEP Theory Group at Argonne National Laboratory, where a part of this work was completed.
Argonne is operated under Contract No. DE-AC02-06CH11357.

\vfill
\pagebreak
\begin{center}{\Large \bf Supplemental Material}\end{center}

\section*{Recoil corrections at tree level}

Consider the matrix element of the weak current at tree level,
\begin{align}
\langle p(p^\prime) | J_W^{+\mu} | n(p) \rangle 
\propto 
\bar{u}^{(p)}(p^\prime) \bigg\{ \gamma^\mu F_1(q^2) + {\iu \over 2 M} \sigma^{\mu\nu} q_\nu F_2(q^2) 
+ \gamma^\mu \gamma_5 F_A(q^2) + {1\over M} q^\mu \gamma_5 F_P(q^2) 
\bigg\} u^{(n)}(p) \,,  
\end{align}
where 
 $q^\mu = p^{\prime \mu} - p^\mu$, $M=(m_n+m_p)/2$, and we will use $\Delta=m_n-m_p$. 
 We have neglected form factors that violate  
 time-reversal invariance isospin symmetry~\cite{Bhattacharya:2011ah}. 
 A straightforward computation for the neutron beta decay process up to $O(1/m_p^2)$, yields\footnote{
$\sum |{\cal M}|^2$ is here interpreted using ``nonrelativistic" normalization of states  i.e., dividing the expression using relativistic normalization by  
$16 m_n E_p E_e E_\nu$.
 }
 \begin{align}\label{eq:recoil}
\sum |{\cal M}|^2 
&\propto 
|C_V|^2 + 3 |C_A|^2 + \left( |C_V|^2 - |C_A|^2 \right) 
{p \cos\theta \over E}
\nl
&\quad 
+ {\Delta \over m_p} 
\bigg\{ 
\left(1-{m^2\over E\Delta}\right) \left( |C_V|^2 +|C_A|^2 \right) 
+ 2\left(1-{2E\over \Delta} + {m^2\over E\Delta}\right)
\,{\rm Re}\left[C_A^*\left(C_V + F_2(0)\right)\right]
\bigg\}
\nl
&\quad
+ {\Delta^2\over m_p^2}
\bigg\{
{m^2\over E\Delta} \left(1-{E\over \Delta} + {p\cos\theta\over \Delta} \right) \, {\rm Re}\left[C_A^* F_P(0)\right] + 
\dots \bigg\}   \,, 
 \end{align}
 where $\cos\theta$ is the angle between the neutrino and electron in the lab frame, and we retain the leading dependence on $F_P$ in the second order correction. The ellipses denote $O(1/m_p^2)$ corrections besides the pseudoscalar form factor which is enhanced by $m_p^2/m_\pi^2$, and which we therefore include in our numerical estimates.  
At tree level we take 
$F_1(0)=C_V= 1$, $F_A(0)=C_A=\lambda$, $F_2(0)=\mu_p-\mu_n-1$, and the pion pole approximation $F_P(0)= (2m_p^2/m_{\pi^+}^2)\lambda$. 
We use central values $\lambda=-1.2764$, $\mu_p-\mu_n-1=3.7059$.

The neutron decay rate is  
\begin{equation}\label{eq:width}
    \Gamma \propto \int \dd\Phi \sum |\mathcal{M}|^{2}~,
\end{equation}
where the phase space (including finite nucleon mass, i.e. beyond the static limit) is  
\begin{equation}
    \dd\Phi \propto \dd E \frac{\dd\cos{\theta}}{2} \qty(\frac{m_{p}}{m_n-E+p\cos{\theta}}) \qty[E_{\nu}(E,\cos\theta)]^{2} p E \,, 
\end{equation}
where 
\begin{equation}
    E_\nu(E,\cos\theta)=\frac{m_n^2-m_p^2+m^2-2 E m_n}{2 \left(m_n-E+p\cos\theta\right)}~.
\end{equation}
Expressing $m_n=m_p+\Delta$, and systematically expanding both phase space and matrix element in powers of $1/m_p$, we have up to $O(1/m_p)$
\begin{align}
    1 + \delta_{\rm recoil} \simeq
 {\int \dd\Phi \sum |\mathcal{M}|^{2} 
 \over \left(\int \dd\Phi \sum |\mathcal{M}|^{2}\right)_{M\to\infty} }
   = 1 -2.068 \times 10^{-3} + (0.005 \times 10^{-3} + \dots )
    \simeq 1 - 2.063 \times 10^{-3}
    \,,
\end{align}
where the successive terms are from the $(\Delta/m_p)^0$, $(\Delta/m_p)^1$ and $(\Delta/m_p)^2$ expansion of \cref{eq:width}. The $(\Delta^2/m_p^2)$ term includes only the contribution from the pseudoscalar form factor which is enhanced by $m_p^2/m_\pi^2$; other $O(1/m_p^2)$ corrections from both the matrix element and the phase space have been dropped.

\vfill 
\pagebreak 

\section*{Radiative recoil}

The leading radiative recoil correction (i.e., including only $\pi$-enhanced terms) is given by 
\begin{equation}
  \delta_{\rm rad.rec.,1}  =  
 {\int \dd\Phi \sum |\mathcal{M}|^{2} \left[ {\pi \alpha \over \beta} \right]
 \over \left(\int \dd\Phi \sum |\mathcal{M}|^{2}\right)_{M\to\infty} }
 = -0.0703 \times 10^{-3} + (0.0048 \times 10^{-3} + \ldots) \,,
\end{equation}
where the successive terms are from the $(\Delta/m_p)^1$ and $(\Delta/m_p)^2$ terms in Eq.~(\ref{eq:width}). As above the $\Delta^2/m_p^2$ correction contains only the contribution from the pseudoscalar form factor. 
We include also the correction accounting for the difference between Coulomb corrections in the neutron and proton rest frame: 
\begin{equation}
    \delta_{\rm rad.rec.,2}  =  
 {\int \dd\Phi \sum |\mathcal{M}|^{2} \left[ {\pi \alpha}
 \left( {1 \over \beta^\prime} - {1\over \beta} \right) \right] 
 \over \left( \int \dd\Phi \sum |\mathcal{M}|^{2}  \right)_{M\to\infty} }
 = - 0.0122 \times 10^{-3} \,,
\end{equation}
For the sum, 
\begin{equation}
    \delta_{\rm rad.rec.} = \delta_{\rm rad.rec.,1} + \delta_{\rm rad.rec., 2} = -0.078 \times 10^{-3}~.
\end{equation}

\bibliography{largepi}

\end{document}